\documentclass[wssquare]{ws-procs961x669} 
\usepackage{ws-toc,ws-multind}
\usepackage{graphicx}	
\usepackage{natbib}
\usepackage{float}

\usepackage{amsmath,amssymb,amsfonts}

\usepackage{natbib}

\begin{document}









\wstoc{For proceedings contributors: Using World Scientific's\\ WS-procs961x669 document class in \LaTeX2e}
{Samik Mitra}

\title{Properties of low angular momentum general relativistic MHD flows around black holes}

\author{Samik Mitra$^*$}
\index{author}{Author, A} 

\address{Department of Physics, Indian Institute of Technology Guwahati,\\
Guwahati, Assam 781039, India\\
$^*$E-mail: m.samik@iitg.ac.in}

\begin{abstract}
In this proceeding, we provide a novel approach to study the General Relativistic Magnetohydrodynamic (GRMHD) accretion flows around rotating black holes (BHs). In doing so, we choose a sub-Keplerian distribution of angular momentum of the flow, which is necessary for the accreting matter to reach the event horizon of the BH. Further, we consider the convergent flow to be confined about the disk mid-plane and is threaded by both radial ($b^r$) and toroidal ($b^\phi$) magnetic field components. For simplicity, we neglect any motion along the vertical ($\theta$) direction, maintaining a vertical (hydrostatic) equilibrium about the midplane. With this, we describe the family of multi-trans-magnetosonic accretion solutions around rotating BHs and examine how the conserved magnetic flux ($\Phi$) and BH spin ($a_{\rm k}$) affect the accretion flow properties. Finally, we provide an insight into the thermal emissions from the magnetized disk for a specific set of accretion solutions.

\end{abstract}


\bodymatter

\section{Introduction}\label{aba:sec1}

Recent observations of large-scale magnetic fields \cite[Event Horizon Telescope collboration (EHTC)]{EHT_Mag2021} associated with super-massive black holes (SMBHs) confront the prospect that the magnetic fields could potentially affect the accretion-ejection processes. In a seminal work, Shakura \& Sunyaev \cite{Shakura-Sunyaev1973} predicted that the angular momentum transport in an accretion disk may possibly be driven by magneto-hydrodynamical (MHD) turbulence. However, the underlying physical mechanism was identified almost two decades later by Balbus \& Hawley \cite{Balbus-Hawley1991,Balbus-Hawley1998} in their exceptional work on magneto-rotational instability (MRI). Since magnetic fields are ubiquitous in the universe, accretion disks around BHs are also expected to be threaded by large-scale magnetic fields, which are rooted either from the low-mass companion star or from the interstellar medium \cite{Kogan-Ruzmaikin1974,Kogan-Ruzmaikin1976,Kogan-Lovelace2011}. Based on the disk geometry and dynamics of plasma around BHs, toroidal magnetic fields come out to be the simplest choice \cite{Oda-etal2007,Oda-etal2010,Oda-etal2012}. However, the plunging region is governed by the poloidal magnetic fields \cite{Hawley2001,Kato-etal2004}. Recently, we \cite{Mitra-etal2022,Mitra-Das2024} have shown analytically that for an ideal (MHD) fluid, the toroidal magnetic field dominates in regulating the mid-plane disk dynamics over the radial component around BHs. In the present analysis, we adopt the same model assumptions of our previous work \cite{Mitra-Das2024} and consider the mid-plane to be threaded by radial as well as toroidal magnetic fields around a rotating BH. With this motivation, we discuss the semi-analytic modeling of GRMHD accretion flows around rotating BHs and understand the effect of BH spin, and the underlying spectral energy distributions (SEDs).

Semi-analytic models of accretion flow are generally characterized by the energy and angular momentum ($\lambda$) of the inflowing matter, viscosity, and maybe magnetic field strength or magnetic flux \cite[]{Das2007,Mitra-Das2024}. Needless to mention that, flow with the Keplerian angular momentum ($\lambda_{\rm Kep}$) terminates at the innermost stable circular orbit (ISCO, \cite{Shakura-Sunyaev1973}) and fails to reach the horizon ($r_{\rm h}$). However, the sub-Keplerian ($\lambda \ll \lambda_{\rm Kep}$) matter entering from the outer edge of the disk smoothly connects to the event horizon with supersonic velocity. This confirms the transonic nature of accretion flows around BHs. It is worth mentioning that the observed spectrum of black holes is partially derived from a Keplerian disk, which emits multi-color black body radiation \cite[]{Shakura-Sunyaev1973}, and partly from a power-law component, mainly originating from the quasi-spherical, low angular momentum flow component \cite[]{Abramowicz-etal1988,Chakrabarti-Mandal2006}. Additionally, Proga \& Begelman \cite{Proga-Begelman2003} demonstrated numerically that matter with very low angular momentum plays a crucial role in determining the total mass supply and the accretion rate onto BHs. Also, in a Roche lobe overflow (in XRBs) matter usually matter begins to accrete from the Bondi radius ($\sim 10^6$ gravitational radius), which requires sufficiently large computational support to understand the accretion dynamics. However, in the case of low angular momentum flow, the circularization radius becomes of the order of a few hundred gravitational radii ($r_{\rm g}$) of the accretor \cite[]{Frank-etal2002,Moscibrodzka-etal2007}. In this context, few global GRMHD simulations are also seeking motivations from the low angular momentum studies \cite[]{Ressler-etal2020a,Ressler-etal2020,Kwan-etal2023,Olivares-etal2023,Dihingia-Mizuno2024,Galishnikova-etal2024}. Therefore, the theoretical modeling of low angular momentum flows plays an important role in understanding the large-scale feeding process.

\section{GRMHD flow: Formalism and underlying assumptions} \label{EQN}

In this proceeding, we aim to study the magnetized, low angular momentum hot accretion flows around a stationary, axisymmetric space-time. In Boyer-Lindquist coordinates, the standard form line element of such space-time is as follows \citep{Kerr1963}:
\begin{eqnarray}\label{rotmetric}
ds^2 & =  - \left( 1- \frac{2r}{\Sigma} \right) dt^2  - \frac{4a_{\rm k}r}{\Sigma  } \sin^2 \theta dt \; d\phi & +
\frac{\Sigma}{\Delta}dr^2  +  \Sigma d \theta^2+ \nonumber
\\ & &\left[r^2+ a_{\rm k}^2 +
\frac{2 r a_{\rm k}^2 }{\Sigma} \sin^2 \theta
\right] \sin^2 \theta d\phi^2,
\end{eqnarray}
where $a_{\rm k}$ is the BH spin and
\begin{eqnarray*}
\Sigma = r^2 + a_{\rm k}^2 \cos^2\theta;\;\;\;\;\;  \Delta = r^2 + a_{\rm k}^2 - 2r.
\end{eqnarray*} 
For the analysis, we express length $r$ and time $t$ in terms of $r_{\rm g}$ and $r_{\rm g}/c$ where $r_{\rm g}=GM_{\rm BH}/c^2$. Here, $G$ is the gravitational constant, $c$ is the speed of light, and $M_{\rm BH}$ refers to the mass of BH. Using this unit system, we write the governing GRMHD equations \citep[]{Lichnerowicz1970, Anile1990} as,
\begin{equation}
\nabla_\mu (\rho u^\mu) = 0; \hspace{0.5cm} \nabla_\mu T^{\mu\nu}=0; \hspace{0.5cm} \nabla_\mu {}^* F^{\mu\nu}=0,
\end{equation} where $\rho$ is the mass density, $u^\mu$ is the four velocity, $T^{\mu \nu}$ is the energy-momentum tensor, and ${}^*F^{\mu\nu}$ is the dual of Faraday's electromagnetic tensor. We consider that the accreting fluid has infinite conductivity, which allows the magnetic field lines to remain frozen into the accreting plasma following ideal GRMHD conditions, \textit{i.e.,} $u_\mu b^\mu =0$. As a consequence the electric field ($e^\mu$) in fluid frame goes to zero. In a magnetized flow, the energy-momentum tensor is given by \citep{Abramowicz-Fragile2013},
\begin{equation}
\begin{aligned}
T^{\mu}_\nu &=(T^\mu_\nu)_{\rm Fluid} + (T^\mu_\nu)_{\rm Maxwell}\\
&= \{ (e+p_{\rm gas}) u^\mu u_\nu + \delta^{\mu}_{\nu}  p_{\rm gas} \} + \{ \delta^\mu_\nu \frac{b^2}{2} + b^2 u^\mu u_\nu- b^\mu b_\nu \}.
\end{aligned}
\end{equation}
Here, $p_{\rm gas}$ is the gas pressure, $e$ is the internal energy, $b^\mu$ refers to the four magnetic fields in the comoving frame, and $b^2=b_\mu b^\mu$. Note that, $\delta^\mu_\nu = g^{\mu\alpha} g_{\alpha\nu}$ is the contraction of the covariant and contravariant metric components.

\subsection{Globally conserved quantities MHD flows}

Considering the convergent flow ($u^r < 0$) to remain around the disk mid-plane, \textit{i.e.,} $\theta=\pi/2$, we choose the polar component of the four-velocity as $u^\theta \sim 0$. Furthermore, for simplicity we keep $b^\theta=0$, making the radial ($b^r$) and toroidal ($b^\phi$) field components independent. With this, we aim to examine the radial behavior of the advective, axisymmetric ($\partial_\phi \rightarrow 0$) flow in the steady-state ($\partial_t \rightarrow 0$).

Firstly, from the particle number conservation equation, we get
\begin{equation}
	\sqrt{-g} \rho u^r =  C_{\mathcal{M}},
\end{equation}
where $C_{\mathcal{M}}$ is a constant measure of the mass flux and the determinant $\sqrt{-g}=r^2$ for the Kerr-BH in $\theta=\pi/2$ limit. Needless to mention that, the Kerr metric is stationary and axisymmetric, it is associated with two Killing vector fields. As the fluid is assumed to obey the symmetries of the chosen space-time, the revised energy-momentum conservation merged with the Killing equation takes the following form $\nabla_\mu(T^{\mu}_\nu \xi^\nu)=0$, where $\xi^\nu$ refers to the generic killing vectors. With this, we obtain the globally conserved specific energy flux ($\mathcal{E}$) for $\nu=t$ as
\begin{equation}
	-\frac{\sqrt{-g} T^r_t}{C_{\mathcal{M}}} = \mathcal{E},
\end{equation}
and the conserved specific angular momentum flux ($\mathcal{L}$) is obtained for $\nu=\phi$ as
\begin{equation}
	\frac{\sqrt{-g} T^r_\phi}{C_{\mathcal{M}}} = \mathcal{L}.
\end{equation}

Additionally, the no-monopole constrain \cite[]{Porth-etal2019} implies,
\begin{equation}
-\sqrt{-g}{}\, {}^*F^{rt}=\Phi,
\end{equation}
where $ {}^*F^{rt} =u^t b^r - u^r b^t$ and $\Phi$ is referred to the measure of radial magnetic flux. Next, the $\phi$-component of source-free Maxwell's equation is written as \cite[]{McKinney-Gammie2004},
\begin{equation}
\sqrt{-g}{}{}^*F^{r\phi}=F,
\end{equation}
where ${}^*F^{r\phi}=u^r b^\phi - u^\phi b^r$. Equation (8) is known as the relativistic iso-rotation law and we define $F$ as the iso-rotation parameter. Finally, we derive the $r$-component of the Navier-Stokes equation by projecting the energy-momentum conservation equation radially in the fluid frame \cite[]{Mitra-etal2022}, 
\begin{equation}
\begin{aligned}
       (g^{r\nu}+u^ru^\nu) \nabla_\nu p_{\rm tot} + \rho h_{\rm tot} u^\nu \nabla_\nu u^\alpha - \nabla_\nu (b^r b^\nu) - u^r u_\mu \nabla_\nu (b^\mu b^\nu)=0
\end{aligned}
\end{equation}
where the projection operator is $\gamma^r_\mu = \delta^r_\mu + u^r u_\mu$, total pressure is $p_{\rm tot}=p_{\rm gas} + p_{\rm mag}$, magnetic pressure is $p_{\rm mag}=b^2/2$, and the total enthalpy is $h_{\rm tot}= (e+p_{\rm gas})/\rho + b^2/\rho$. 

Considering hydrostatic equilibrium in the vertical direction for the magnetized disk, we calculate the local half-thickness of the disk ($H$) following \cite{Riffert-Herold1995, Peitz-Appl1997}, which is given by,
\begin{equation}
      H^2 = \frac{p_{\rm gas} r^3}{\rho \mathcal F}, \hspace{0.5cm} \mathcal{F}=\gamma_\phi^2 \frac{(r^2+a_{\rm k}^2)^2 + 2 \Delta a_{\rm k}^2}{(r^2+a_{\rm k}^2)^2 - 2 \Delta a_{\rm k}^2},
\end{equation}
where $\gamma_\phi$ $(=1/\sqrt{1-v_\phi^2})$ is the Lorentz factor, $\lambda$ $(=-u_\phi/u_t)$ is the specific angular momentum of the flow, and $\Omega$ $(=u^\phi/u^t)$ is the angular velocity of the fluid \citep{Dihingia-etal2018, Mitra-etal2022}. According to our choice, the magnetic pressure along $\theta-$ direction is zero $b_\theta b^\theta \sim 0$, therefore we consider the hydrostatic balance to be maintained only via the gas pressure. In the next, we follow \cite{Lu1985} to describe the three components of fluid velocities in the corotating frame as $v_\phi^2 = u^\phi u_\phi/(-u^t u_t)$, $v_r^2 = u^r u_r/(-u^t u_t)$, and $v_\theta^2 = u^\theta u_\theta/(-u^t u_t)$ where $v_\theta=0$ as $u^\theta \sim 0$ in the disk mid-plane. Upon integrating Eq. 4, we obtain the globally conserved mass-accretion rate in the comoving frame as follows,
\begin{equation}
	\dot{M} = -4\pi \rho v \gamma_v H \sqrt{\Delta},
\end{equation} where $v$ $(= \gamma_\phi v_r)$ is the flow velocity and $\gamma_v=1/\sqrt{1-v^2}$. In this work, we scale the accretion rate as $\dot{m}=\dot{M}/\dot{M}_{\rm Edd}$, where $\dot{M}_{\rm Edd}$ is the Eddington accretion rate ($\dot{M}_{\rm Edd} = 1.4 \times 10^{18} \big(\frac{M_{\rm BH}}{M_\odot}\big)$ g s$^{-1}$; here $M_\odot$ is the solar mass). Using ideal MHD condition $u_\mu b^\mu=0$ and Eqs. (7-8), we express $b^{r,\phi}$ in terms of $\Phi$ and $F$ as
\begin{equation}
b^r = - \frac{\gamma_\phi^2 (\Phi + F \lambda)}{u^t \mathcal{A}},\;\;\;
b^\phi = \frac{F v^2 - \gamma_\phi^2 (F + \Phi \Omega)}{u^r \mathcal{A}},
\end{equation}
where $\mathcal{A}=r^2 (v^2 - 1)$.
With these transformations, we analyze the magnetized accretion flow using the global constants ($\Phi, F$). So far, we have five equations 

\subsection{Equation of state}
To close the dynamical equations (\textit{i.e.,} Eqs. 5-9 and Eq. 11), we use the relativistic equation of state (REoS; see reference \cite[]{Chattopadhyay-Ryu2009}), which is given by
\begin{equation}
	e = \frac{\rho f}{\big(1+\frac{m_p}{m_e}\big)},
\end{equation}
where $m_e$ and $m_p$ are the masses of electrons and protons. The quantity $f$ is expressed in terms of dimensionless temperature ($\Theta = k_{\rm B} T/m_e c^2$, $k_{\rm B}$ is the Boltzmann constant) as
\begin{equation}
f = \Bigl\{1+ \Theta \bigg(\frac{9\Theta+3}{3\Theta+2}\bigg)\Bigl\} + \Bigl\{\frac{m_p}{m_e}+\Theta \bigg(\frac{9\Theta m_e+3m_p}{3\Theta m_e+2m_p}\bigg)\Bigl\}.
\end{equation}
With this, we define the polytropic index as $N=(1/2) (df/d\Theta)$ and adiabatic index as $\Gamma(r) = 1 + 1/N$. The choice of such EoS is important to explain the trans-relativistic nature of the accretion flows, where $\Gamma$ reaches $4/3$ close to BH and far from the central object $\Gamma \sim 5/3$. Hence, it is obvious to have a radially varying adiabatic index which is self-consistently controlled by the temperature of the disk.

However, it is noteworthy that the MHD flow has several characteristic wavespeeds other than the sound speed. Usually, the characteristic wave speeds for magnetized flows come from the Alfv\'en and magneto-sonic waves, respectively. Following reference \cite{Gammie-etal2003}, we define the Alfv\'en speed as $C_a^2 = b_\mu b^\mu/(\rho h_{\rm tot})$, and the fast-magnetosonic speed as $C_{\rm f}^2= C_{\rm s}^2 + C_a^2 - C_{\rm s}^2 C_a^2$, where the relativistic sound speed is given by $C^2_{\rm s}=\Gamma p_{\rm gas}/(e + p_{\rm gas})$. Moreover, we define the magnetosonic Mach number as $M=v/C_{\rm f}$.

\subsection{Critical point analysis} \label{Method}

We combine equations (5-9, 11) and obtain the following three coupled non-linear differential equations as
\begin{itemize}
\item[(a)] the radial momentum equation:
\begin{equation}
\begin{aligned}
R_0 + R_v \frac{dv}{dr} + R_\Theta \frac{d\Theta}{dr} + R_\lambda \frac{d\lambda}{dr} &=0,
\end{aligned}
\end{equation}
\item[(b)] the azimuthal momentum equation:
\begin{equation}
\begin{aligned}
\mathcal{L}_0 + \mathcal{L}_v \frac{dv}{dr} + \mathcal{L}_\Theta \frac{d\Theta}{dr} + \mathcal{L}_\lambda \frac{d\lambda}{dr} &=0,
\end{aligned}
\end{equation}
\item[(c)] the energy equation:
\begin{equation}
\begin{aligned}
\mathcal{E}_0 + \mathcal{E}_v \frac{dv}{dr} + \mathcal{E}_\Theta \frac{d\Theta}{dr} + \mathcal{E}_\lambda \frac{d\lambda}{dr} &=0.
\end{aligned}
\end{equation}
\end{itemize}
The coefficients in Eqs. (15-17) are obtained in the same way as described in reference \cite[]{Mitra-Das2024}. We recall that, the information of magnetic field is already supplied in these equations following Eq. 7 and Eq. 8.

In the next, using Eqs. (15-17) we obtain the wind equation as,
\begin{equation}
\frac{dv}{dr} = \frac{\mathcal{N}(r,a_{\rm k},v,\Theta,\lambda,\Phi,F)}{\mathcal{D}(r,a_{\rm k},v,\Theta,\lambda,\Phi,F)}.
\end{equation}
	
Similarly, the angular momentum gradient takes the form 
\begin{equation}
\frac{d\lambda}{dr} = \frac{\mathcal{L}_{\Theta} \mathcal{E}_{0} - \mathcal{L}_{0} \mathcal{E}_{\Theta}}{\mathcal{L}_{\lambda} \mathcal{E}_{\Theta} - \mathcal{L}_{\Theta} \mathcal{E}_{\lambda}} + \frac{(\mathcal{L}_{\Theta} \mathcal{E}_{v} - \mathcal{L}_{v} \mathcal{E}_{\Theta})}{\mathcal{L}_{\lambda} \mathcal{E}_{\Theta} - \mathcal{L}_{\Theta} \mathcal{E}_{\lambda}} \frac{dv}{dr},
\end{equation}
	
and the temperature gradient has the form

\begin{equation}
\frac{d\Theta}{dr} = \frac{\mathcal{L}_{\lambda} \mathcal{E}_{0} - \mathcal{L}_{0} \mathcal{E}_{\lambda}}{\mathcal{L}_{\Theta} \mathcal{E}_{\lambda} - \mathcal{L}_{\lambda} \mathcal{E}_{\Theta}} + \frac{(\mathcal{L}_{\lambda} \mathcal{E}_{v} - \mathcal{L}_{v} \mathcal{E}_{\lambda})}{\mathcal{L}_{\Theta} \mathcal{E}_{\lambda} - \mathcal{L}_{\lambda} \mathcal{E}_{\Theta}} \frac{dv}{dr}.
\end{equation}
To obtain accretion solutions, we numerically solve the equations (18-20) for the set of input parameters $(a_{\rm k},\mathcal{E},\mathcal{L},\Phi, F)$. For more details, see the Appendix of reference \cite{Mitra-Das2024}.

In general, the accreting matter begins its journey from the outer edge of the disk with subsonic velocity ($v << 1$) and descends into the BH supersonically ($v\sim 1$) to fulfill the inner boundary conditions imposed by the horizon. Therefore, the flow must change its sonic state at least once, if not more, while passing through the critical point $r_c$.  At the critical point, the wind equation (Eq. 18) takes an indeterminate form \textit{i.e.,} $(dv/dr)|_{r_{\rm c}} = 0/0$, which corresponds to the critical point conditions $\mathcal{N}=\mathcal{D}=0$. However, in reality, the convergent flow remains smooth along the streamlines even while passing through $r_{\rm c}$. Hence, the velocity gradient must be real and finite everywhere. We, therefore, imply l'H$\hat{{\rm o}}$pital rule on Eq. (18) to evaluate the velocity gradient at $r_{\rm c}$. Accordingly, we obtain two unique values of $(dv/dr)|_{r_{\rm c}}$; one of them relates to accretion ($(dv/dr)|_{r_{\rm c}}<0$), and the other one is to the wind, $(dv/dr)|_{r_{\rm c}}<0$. When both $(dv/dr)|_{r_{\rm c}}$ values are real and opposite in sign, saddle or `X'-type critical points form \cite[]{Das2007,Das-etal2022,Mitra-etal2022,Mitra-etal2023, Nazari-etal2024}. Such points have special significance, as transonic solutions can only pass through these points before falling into the BH. Depending on the input parameters, when `X'-type critical points form close to the horizon, we call them inner ($r_{\rm in}$) critical points, whereas the outer ones ($r_{\rm out}$) form far away from the horizon. Low angular momentum accretion flows around BHs often support the existence of multiple critical points (MCP) for a specific range of input parameters $(a_{\rm k},\mathcal{E},\mathcal{L},\Phi, F)$ and such flows are potentially viable to harbor shock waves \citep[and references therein]{Fukue1987, Chakrabarti1989, Das-Chakrabarti2004, Das2007, Dihingia-etal2019, Das-etal2022, Mitra-Das2024}. Hence, in the subsequent section, we study the shock-induced magnetized flows around BHs.

\section{Shock-induced advective GRMHD flow}

Depending on the input parameters ($a_{\rm k},\mathcal{E},\mathcal{L},\Phi,F$), flow becomes supersonic after passing through the outer saddle-type critical point ($r_{\rm out}$), and connects smoothly to the horizon ($r_{\rm h}$). However, as the supersonic flow travels toward the BH, it begins to feel the centrifugal repulsion and momentarily slows down. Thus, the inflowing matter accumulates in the vicinity of the BH, and the ``virtual'' barrier forms. Such a centrifugal barrier cannot hold the accumulation of matter indefinitely and triggers all of a sudden a discontinuous transition of the flow variables \citep{Takahashi-etal2002, Das-Chakrabarti2007,Fukumura-etal2007,Sarkar-Das2016, Sarkar-etal2018, Dihingia-etal2019, Dihingia-etal2020}. In the astrophysical scenario, such transitions are denoted as shock \citep{Fukue1987, Chakrabarti1989, Frank-etal2002}. 

\subsection{Jump conditions for shocked GRMHD flow:}\label{SHC}

Over the years, \cite{Lichnerowicz1970, Appl-Camenzind1988, Takahashi-etal2006, Fukumura-etal2007} formulated the jump conditions for the MHD flows. We adopt those sets of jump conditions below, 
\begin{itemize}
\item [(a)] Mass flux conservation, $\bigg[\rho u^r\bigg]=0,$
\item [(b)] Energy flux conservation, $\bigg[T^{rt}\bigg]=0,$
\item [(c)] Angular momentum flux conservation, $\bigg[T^{r\phi}\bigg]=0,$ 
\item [(d)] Radial magnetic flux conservation, $\bigg[{}^*F^{r t}\bigg]=0$,
\item [(e)] Iso-rotation conservation, $\bigg[{}^*F^{r\phi}\bigg]=0$,
\item [(f)] Pressure balance condition, $\bigg[T^{rr}\bigg]=0.$
\end{itemize}
Here, the square bracket `$[Z] \equiv Z_+ - Z_-$' represents the change of the certain quantity $Z$ across the shock front, \textit{i.e.,} within the pre-shock (`$-$') and post-shock (`$+$') region. Overall, we use the above six conditions (a-f) to find the shock location ($r_{\rm sh}$).

Following the aforementioned conditions, the supersonic flow jumps to the post-shock branch and becomes subsonic. As a result, the kinetic and electromagnetic energies of pre-shock flow are converted into thermal energy, forming the post-shock corona (PSC). Usually, PSC gets hot and dense due to the shock compression. As the flow moves further inwards, the accreting material acquires more radial velocity and becomes supersonic after passing through the inner critical point ($r_{\rm in}$). Hence, we obtain shock-induced trans-magnetosonic accretion flows.

\section{Results}

We investigate the dynamical structure of shock-induced trans-magnetosonic accretion solutions around a Kerr-BH for a set of model parameters, namely $\mathcal{E}$, $\mathcal{L}$, $\Phi$ and $F$, respectively. In doing so, we want to examine the effects of the radial magnetic flux ($\Phi$) and iso-rotation parameter $(F)$ on the GRMHD solutions. Given the small magnitude of the dimensionless radial magnetic flux and the iso-rotation parameter, we denote them as $\Phi = \Phi_{13} \times 10^{-13}$ and $F = F_{15} \times 10^{-15}$, maintaining the notation $\Phi_{13}$ and $F_{15}$ to signify the magnetic flux and iso-rotation parameter values. Moreover, in this work, we choose $M_{\rm BH}=10M_\odot$ and $\dot{m}=0.001$ as fiducial values unless stated otherwise.  

\begin{figure}[ht!]
\centering
\includegraphics[width=\columnwidth]{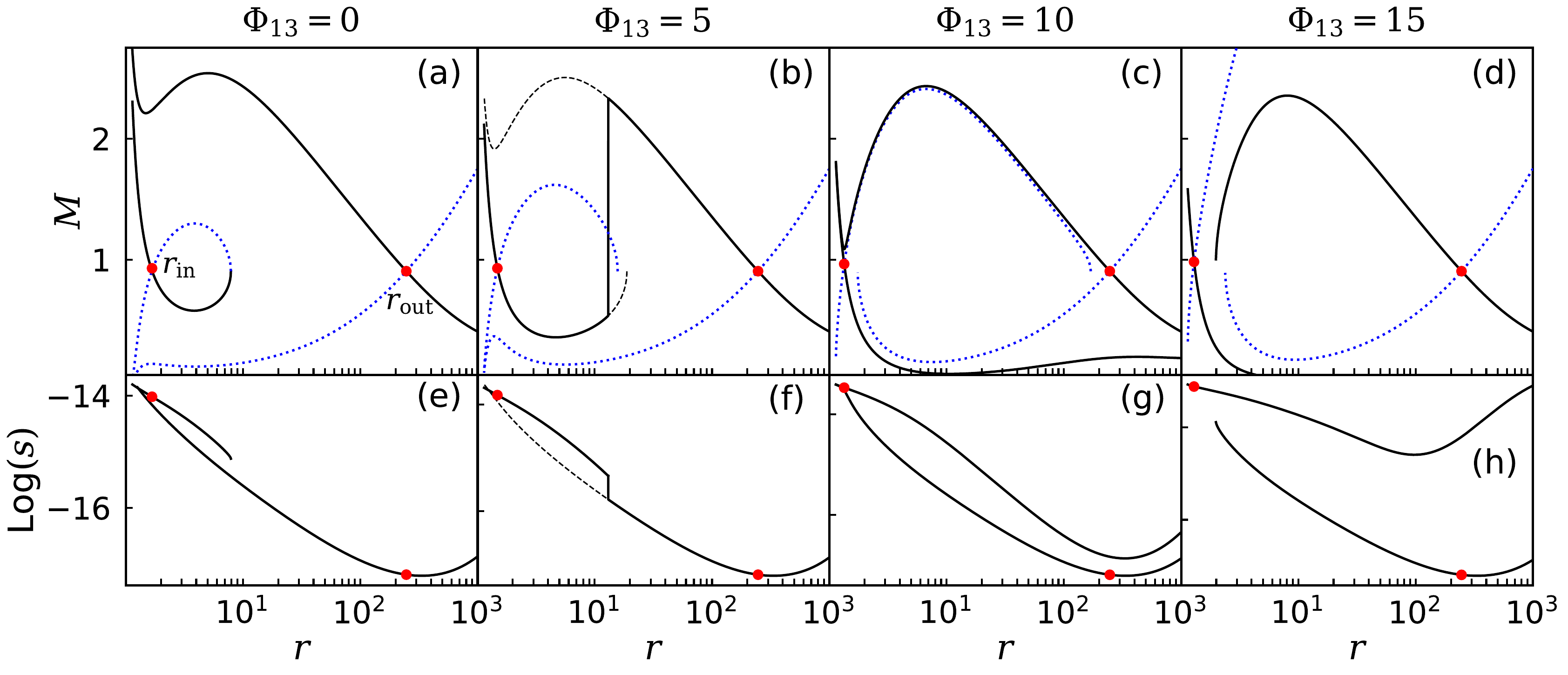}
\caption{Upper panels (a-d): Variation of fast-magnetosonic Mach number ($M =v/\sqrt{C_{\rm s}^2+C_{\rm a}^2 - C_{\rm s}^2 C_{\rm a}^2} $) as a function of logarithmic radial distance ($r$). We vary the radial magnetic flux to check the dependency of $\Phi_{13}$ on accretion solutions. In the lower panels (e-h), we depict the entropy function. }
\label{fig:1}
\end{figure}

In the upper panels of Fig. \ref{fig:1}, we depict the variation of fast magnetosonic Mach number with logarithmic radial distance ($r$). To begin with, we fix the input parameters $(\mathcal{E,L},a_{\rm k},F_{15})=(1.001,1.925,0.99,5)$ and vary the radial magnetic flux. In Fig. \ref{fig:1}a, for $\Phi_{13}=0$, we observe the existence of inner ($r_{\rm in}$) and outer ($r_{\rm out}$) critical points (see the filled circles). However, the accretion solution passing through $r_{\rm in}=1.325$ fails to connect the horizon to the outer edge of the disk. But, the transonic solution passing through $r_{\rm out}=234.556$ smoothly enters into the horizon from the outer edge of the disk, $r_{\rm edge}=1000$. Note that in the corresponding lower panel, we plot the radial variation of the entropy function ($s=p_{\rm tot}/\rho^{\Gamma-1}$). It shows that the entropy content of solutions passing through $r_{\rm in}$ is higher than the outer critical point passing solution. 

As we increase the flux to $\Phi_{13}=5$, we notice in Fig. 1b, that the supersonic solution passing through $r_{\rm out}$ no longer enters directly into the horizon. Instead, it follows a path whether entropy content is higher. Hence, the supersonic flow jumps into the subsonic branch via shock. Eventually, the subsonic flow gains radial velocity and passes through $r_{\rm in}$ by becoming supersonic. Finally, the flow falls into the BH, and we obtain a shock-induced accretion solution. Needless to mention that the shock-induced global trans-magnetosonic solution contains higher entropy than the shock-free solution (dashed curve) \cite[]{Fukue1987,Chakrabarti1989,Chakrabarti-Titurchuk1995}. 

Further, we report a special class of GRMHD solutions for $\Phi_{13} = 10$ where both the accretion solutions passes through $r_{\rm in}$ and $r_{\rm out}$ (see panel (c)). These solutions connect the $r_{\rm edge}$ to the horizon. Hence, there is a degeneracy present. Here, the entropy saves us, since the flow passing through $r_{\rm in}$ has higher entropy content than in the solutions passing through $r_{\rm out}$. Next, we increase $\Phi_{13}=15$ and topology of accretion solution changes and becomes `W'-type as discussed in \cite[]{Dihingia-etal2018}.

\subsection{Effect of prograde and retrograde BHs onto the accretion flow structure}

\begin{figure}[ht!]
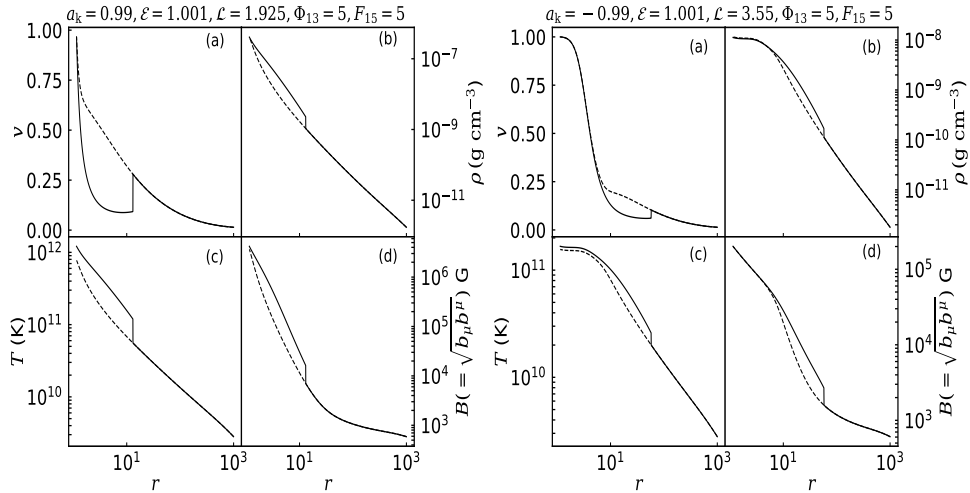

\centering
\includegraphics[height=6.5cm,width=6.25cm]{Shock_variables}
\includegraphics[height=6.5cm,width=6.25cm]{Shock_pm99}
\caption{Comparative analysis between the flow variables around a prograde (left panel) BH with $a_{\rm k}=0.99$ and a retrograde BH with $a_{\rm k}=-0.99$ (in right panel). The input parameters in both cases are mentioned at the top of the figures.}
\end{figure}

According to GR, spin-orbit coupling is unavoidable around a spinning BH. This is observed when we try to write down the effective potential around a Kerr BH (see \cite[]{Dihingia-etal2018}). This appears due to the frame-dragging effect produced mainly by the rotation of the BH. As a result of such a dragging effect, the convergent matter is forced more toward the azimuthal direction instead of radial infall. Hence, we notice the growth of frictional force within the disk, which rapidly grows the flow's temperature. Consequently, we notice a rapid increase in the net magnetic field due to the rapid growth of toroidal components around a prograde BH. 

On the contrary, when the BH is spinning oppositely with the rotation of the disk, we notice a reduced frame-dragging effect. In such a scenario, the central BH is referred to a retrograde BH. First, we notice that the location of critical points shifts more outwards than in the prograde case. Secondly, due to the reduction of the effective angular momentum in the system, we notice a lower temperature and magnetic field strength than the results in the prograde case. 

Additionally, for the retrograde case, the effective angular momentum is reduced, and high angular momentum is required to sustain shocks, which is the opposite in the case of prograde BHs.

\begin{figure}[ht!]
\centering
\includegraphics[height=6.5cm,width=6.25cm]{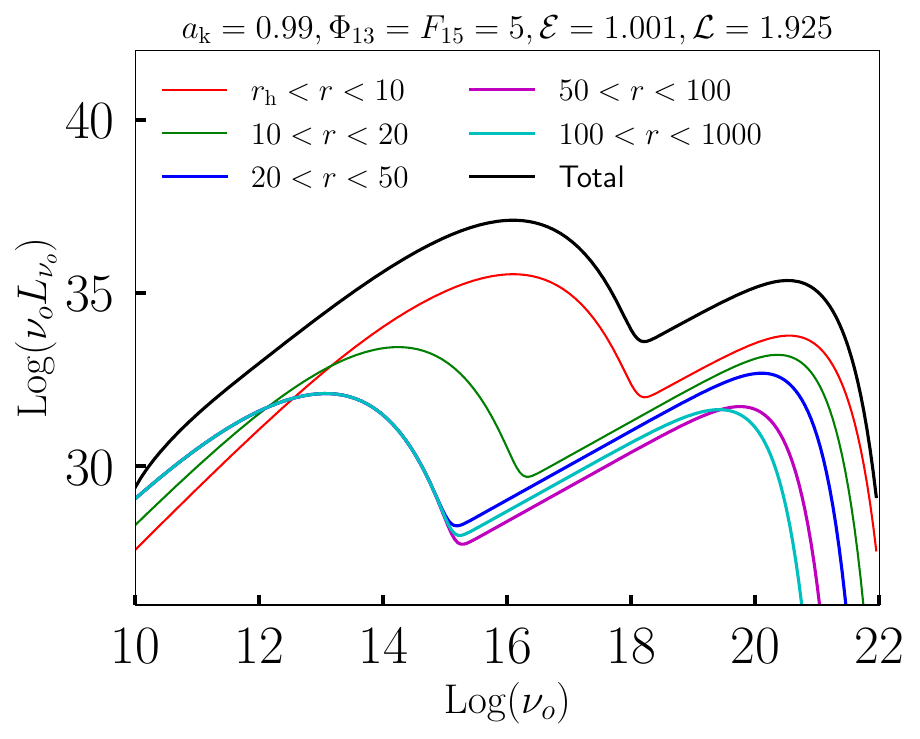}
\includegraphics[height=6.5cm,width=6.25cm]{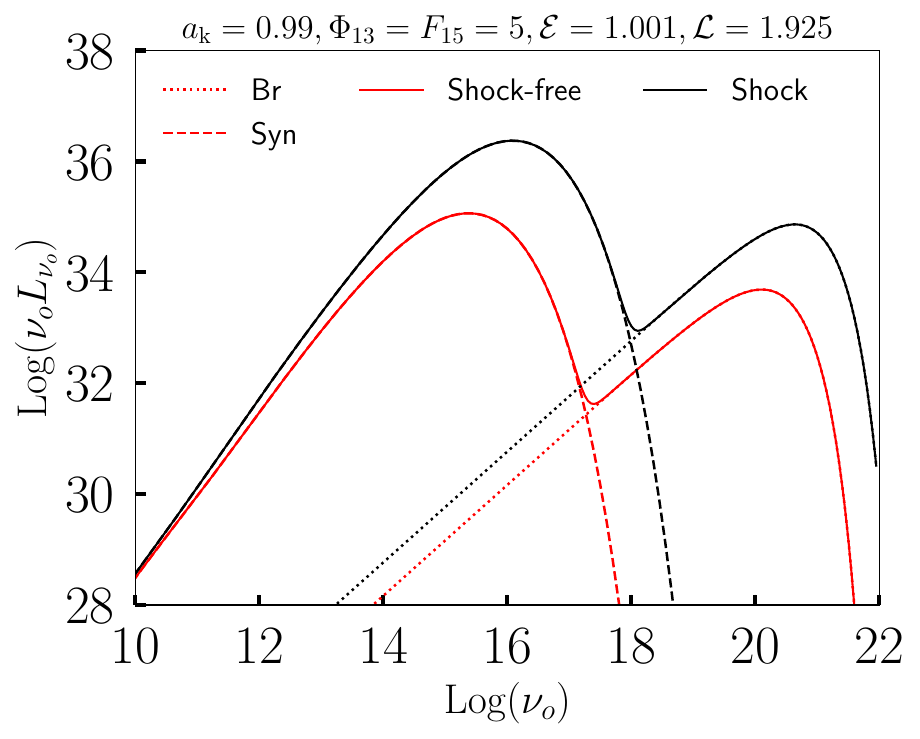}
\caption{A pictorial depiction of the spectral energy distribution (SED) incorporating the free-free emission (Bremsstrahlung, dotted) and thermal synchrotron (dashed) process. Left panel: Composite SEDs from different parts of the accretion disk. Right panel: Comparison of SED between a shock and shock-free solution.}
\end{figure}
\section{Discussion and Conclusion}
In this study, we examine the global structure of shock-induced, magnetized advective accretion flows around Kerr black holes (BHs) in a steady state. Unlike the work of \cite[]{Mitra-etal2022}, we emphasize the importance of magnetic flux profiles (\(\Phi\) and \(F\)) in regulating the dynamics of the accretion disk, rather than focusing solely on local magnetic fields (\(b^r\) and \(b^\phi\)). By adopting a relativistic equation of state, we solve the governing equations of GRMHD flow to derive global trans-magnetosonic accretion solutions that exhibit fast MHD wave speeds.

Our findings reveal that, depending on the chosen input parameters (\(\mathcal{E}\), \(\mathcal{L}\), \(\Phi\), and \(F\)), the flow can possess multiple critical points (MCP). These multi-transonic accretion solutions hold significant implications in astrophysics, as they may harbour shock waves, which are crucial for understanding the hard X-ray radiations coming from Galactic BHs (GBHs). The dynamics of these shocks are strongly influenced by several factors, including the black hole spin (\(a_{\rm k}\)), the energy (\(\mathcal{E}\)), the angular momentum flux (\(\mathcal{L}\)) of the flow, the radial magnetic flux (\(\Phi\)), and the iso-rotation parameter (\(F\)). We specially focus on the effect of prograde and retrograde BHs and the frame-dragging effect shows the importance

However, it is interesting to understand the thermal radiations emanating from the disks. Hence, I have calculated the SED coming from the free-free emission and the synchrotron radiation in the presence of magnetic fields. To calculate the emissivity functions we follow \cite[]{2000, Sarkar-Chattopadhyay2019,Sarkar-etal2020, Patra-etal2024}. In Fig. 3, we report that the inner part ($r_{\rm h} < r < 10$) of the disk mostly contribute to the SED, where as the outer region from the disk is not producing radiations effectively. Hence, again it is proved the importance of low angular momentum flow as the sub-Keplerian matter can enter inside ISCO eventually falling onto the BH.

Finally, we compare the SED in shock-induced and shock-free solutions. We see that the shock-induced solutions produce much higher SEDs than the shock-free solutions. This happens because the temperature of a shock solution is larger than that of a shock-free solution. A detailed analysis of the SED will be reported elsewhere.

\section*{Acknowledgement}

I thank the Marcel Grossman team for selecting my work for oral presentation and Prof. Banibrata Mukhopadhyay for reviewing the proceeding as the Editor. SM is highly indebted to Prof. Santabrata Das for providing encouragement to apply for the MG17. SM also thanks the prestigious Prime Minister's Research Fellowship for the financial support.

\bibliographystyle{ws-procs961x669}

\bibliography{ws-pro-sample}





\end{document}